# EVIDENCE FOR CRITICAL INTERNAL HEAT VALUES DURING SIGNIFICANT GEOPHYSICAL TRANSITIONS IN THE INNER SOLAR SYSTEM PLANETARY BODIES IN ASSOCIATION WITH VOLCANISM

Varnana.M.Kumar [1], T.E.Girish [2*], Thara.N.Sathyan [1], Biju Longhinos[3] Anjana AV Panicker[4] and Binoy.J[1]

[1]Department of Physics, Government College for Women, Thiruvananthapuram, Kerala, India 695014

[2]Department of Physics, University College, Thiruvananthapuram, Kerala, India 695034

[3]Department of Civil Engineering, College of Engineering, Thiruvananthapuram, Kerala, India 695016

[4]Department of Geology, University College, Thiruvananthapuram, Kerala, India 695034

*Corresponding Author E mail : tegirish5@yahoo.co.in

We found evidence for critical internal heat values during significant geophysical transitions in the inner solar system planetary bodies in association with volcanism. From a simple rocky planet thermal evolution model, we could infer critical surface heat flux values during peak phases (~1.2 W/m$^2$) and cessation phases (~ 0.092 W/m$^2$) of major volcanism in Earth, Moon, Mars, Venus and Mercury. The above phases of volcanism are accompanied by significant geophysical transitions like growth and decay of global planetary magnetic fields which is likely to be related to systematic changes in core-mantle boundary heat flux values. The above results suggest the that planets are possibly self-organised physical systems with strong core-mantle-crust coupling . The present study will have implications on the search for habitable extrasolar planets.

Key words : critical internal heat, rocky planets, solar system, volcanism, geophysical transitions, CMB heat flux, magnetic fields, self-organisation, extra solar planets,

## 1.Introduction

Thermal history of a planet is related to its internal heat evolution which controls diversified geophysical phenomena such as volcanism, crust formation and changes, planetary magnetic fields, climate etc [1-3] Planetary volcanism is now known to be among the important geophysical phenomena considered for judging the habitability of extra solar planets [4] .In this context understanding the geological time evolution of volcanism in the inner solar system is very much relevant [5] We wish to pose some important questions in this connection: (i) Is there are any patterns and similarity in the time evolution of volcanism and internal heat observed for different rocky planetary bodies in the inner solar system (Earth, Moon, Venus, Mars and Mercury)? (ii) If this is so what is physical basis for the same? (iii) How the results of above investigations will affect the future geophysical evolution of Earth and will be useful for understanding extrasolar planetary volcanism. (iv) In spite of the apparent diversity and complexity in the geophysical characteristics can we consider planets also as self-organised physical systems? In this paper we will try to find answers to some of these questions using a simple model and best available geophysical observations. This work is an update of our previous studies in this direction ( 6-7).

We found evidence for critical internal heat values during significant geophysical transitions in the inner solar system planetary bodies in association with volcanism. Using the rocky planet thermal evolution model, we could infer almost identical surface heat flux values during peak phases (~1.2 W/m2) and cessation phases (~ 0.092 W/m2) of major volcanism in Earth, Moon, Mars, Venus and Mercury. The above phases of volcanism are accompanied by significant geophysical transitions like growth and decay of global planetary magnetic fields (Mars and Moon) , formation of core ( Earth and Moon) submergence of major water bodies (Mars) etc. Some of these geophysical phase transitions are found to be related to changes in critical core-mantle boundary (CMB) heat flux values [4] The above results suggest that planets are also possibly self-organised physical systems as suggested in some previous studies. The implications our model on the search for habitable extrasolar planets will be also discussed.

## 2. Critical internal heat values and distinct geophysical phenomena during the peak phase of volcanism in Earth, Mars, Moon, Venus and Mercury

In Table 1 we have shown the geological time interval of occurrence of peak volcanism in different planetary bodies in the inner solar system. Basic parameters given in this Table is adopted from our previous paper [7].

The surface heat flux of each planetary body in our model [6] is assumed to evolve with time as:

$$S(t) = S_o \exp(-\lambda t) \qquad (1)$$

Here

$S_o$ is the surface heat flux at time of formation of the planetary body

$S(t)$ is the surface heat flux at time t reckoned in Gyrs

$\lambda$ is the decay constant for Chondrite composition

Assuming the current S values of different planetary bodies we can calculate 'So' for each planetary body from equation (1) and can study its thermal evolution with geological time Following the details given our earlier studies [6-7] we have calculated the 'mean S' corresponding to the time interval of peak volcanism for Earth, Mars, Moon, Venus and Mercury. The results are given in Table 1. Surprisingly the mean S values during peak volcanism epoch for different planetary bodies are almost identical. The average of mean S for all rocky planetary bodies during peak volcanism period in our solar system is found to be

Average Mean S (peak volcanism phase) = $0.12 \pm 0.0109$ W/m$^2$ (2)

The value in (2) can be considered as the critical value of internal heat during the peak phase of volcanism which is common to all rocky planets in our solar system and it is expressed as surface heat flux.

**Table 1: Parameters and Geophysical phenomena associated with the Peak phase of major volcanism in rocky planetary bodies in our solar system**

| Planetary body | Peak phase of major volcanism (Gyrs) | Mean S during peak phase (Wm-2) | VEI during peak phase | Associated geophysical phenomena/References |
|---|---|---|---|---|
| Moon | 3.6-3.9 | 0.1062 | 10-11 | Strong Planetary Magnetic fields [2, 8-9] |
| Mercury | 3.55-4.1 | 0.11 | 10-11 | Magnetic field generation started [10] |
| Mars | 3.7-4.1 | 0.12 | 12-13 | Large water bodies and planetary magnetic fields [11] |
| Venus | 0.5 -1 | 0.125 | 9-12 | Intense and widespread volcanism [12] |
| Earth | 0.5 -1 | 0.133 | 10-12 | Snow ball Earth Ice age and inner core formation [13-14 ] |

We could identify distinct geophysical phenomena which has happened during the peak phase of volcanism in different planetary bodies as summarized in Table 1 and allied references are also given.

**3 Critical internal heat values and distinct geophysical phenomena during the cessation phase of major volcanism Earth, Mars, Moon, Venus and Mercury**

Using a simple thermal evolution model for rocky planetary bodies as described in Section 2 and best available observations it is found [6] that major volcanism ceased in Mars, Moon, Mercury and Venus when their inferred surface heat flux is within 10 % of the current surface heat flux value for Earth. These reported results are reproduced in Table 2.

**Table 2: Parameters and distinct geophysical phenomena related to cessation of major volcanism in different rocky planetary objects in the inner solar system (* The values for Earth is calculated from the mean S value during cessation of major volcanism in other planetary objects)**

| Planetary body | Cessation age of major volcanism from current epoch (Gyr) | Inferred S during cessation | Maximum VEI during cessation | Associated geophysical phenomena/References |
|---|---|---|---|---|
| Moon | 3.30 | 0.085 | ---- | Disappearance of lunar magnetosphere [2-3] |
| Mercury | 3.50 | 0.094 | ---- | Global contraction [15] |
| Mars | 3.50 | 0.099 | ---- | Disappearance of water bodies, magnetic fields and major climate/atmospheric changes [11,16] |
| Venus | 0.0025 | 0.091 | 7 | Spatially restricted and low intensity volcanism [17] |
| Earth | 20 Ma* in Geological future | 0.0922* | 7* | Reversibility of habitable conditions ? |

The mean S value during the cessation of major volcanism in Moon, Mars, Mercury and Venus is found to be $0.922 \pm 0.0045$ W/m$^2$ which may be considered as the critical value of surface heat flux for cessation of major volcanism in rocky planetary objects. The surface heat flux of Earth is inferred to decrease to this value in 20 million years from now which can be the cessation age of major volcanism in our planet as shown in Table 2. Distinct geophysical phenomena/transitions in different planetary objects at the time cessation of major volcanism is also shown in Table 2.

**4. On the asymmetry in the ascending and descending phases of major volcanism in rocky planetary objects in the inner solar system**

In Table 3 we have shown our calculations of the total duration (Geological time interval from planetary formation to cessation age of major volcanism), ascending phase (Geological time interval from planetary formation and centre age of peak phase of volcanism) and descending phase (geological time interval from centre age of peak phase and cessation age of major volcanism) in different planetary objects in the inner solar system.

The asymmetry factor in evolution of major volcanism is defined as the ratio of duration of ascending phase to that of descending phase. The asymmetry factor inferred for different planetary bodies in the inner solar system is shown in Table 3 . This factor is found to be maximum for Earth and minimum for Mars.

**Table 3: Asymmetry in duration of ascending and descending phases of major volcanism in inner solar system planets ( * estimated values)**

| Planetary body | Duration of major volcanism (Gyrs) | Centre age of peak phase of volcanism | Duration of ascending phase (AP: Gyrs) | Duration of descending phase (DP: Gyrs) | Asymmetry factor (AP/DP Ratio) |
|---|---|---|---|---|---|
| **Moon** | 1.2 | 3.6 Gyrs | 0.9 | 0.3 | 3 |
| **Mercury** | 1 | 3.82 Gyrs | 0.68 | 0.32 | 2.12 |
| **Mars** | 1 | 3.9 Gyrs | 0.6 | 0.4 | 1.5 |
| **Venus** | 4.4975 | 750 Myrs | 3.75 | 0.75 | 5 |
| **Earth** | 4.52* | 750 Myrs | 3.75 | 0.77* | 4.87 |

## 5 Discussion

Different phases of volcanism in solar system rocky planetary bodies are associated with definite values of internal heat. The peak phases of volcanism in these planetary bodies is associated with a surface heat flux ~ 0.12 W/m2 ( with a standard error of 9 % about this mean value) and cessation phases of volcanism in the same is found to be associated with a surface heat flux of ~0.092 W/m2 ( with a standard error of about 5 % around this mean value) The geological time of occurrence of these volcanic phases are different for different planetary bodies. So, it is a surprise how a planetary body "remembers" to be in a particular geophysical state when its internal heat reaches a given value of surface heat flux. Further distinct geophysical transitions is also inferred to have happened during different phases of rocky planet volcanism in the past as described below. These results provide evidence for self-organised physical behaviour of planets [18]

**Table 4 Geophysical transitions and phenomena associated with different phases of volcanism inferred for Mars, Moon and Earth**

| Planetary Body | Peak phase of volcanism | Declining Phase of volcanism | Cessation phase of volcanism |
|---|---|---|---|
| Mars | Moderate to strong planetary magnetic fields, active dynamo and supercritical CMB heat flows | Moderate to strong planetary magnetic fields and critical CMB heat flows | Cessation or weakening of dynamo and sub-critical CMB heat flows |
| Moon | Strong lunar magnetic fields and Supercritical CMB heat flows | Lunar magnetic fields showing a decreasing trend. But CMB heat flow is still critical | Cessation or significant weakening of dynamo and sub-critical CMB heat flows |
| Earth | Significant enhancement of internal magnetic fields associated with inner core formation Super critical CMB heat flows | Continued operation of geodynamo and critical CMB heat flows | Possible cessation or weakening of Geo dynamo and sub-critical CMB heat flows? |

Significant geophysical phase transitions/phenomena is found to happen during the peak, declining and cessation phases of major volcanism defined in this paper for inner solar system rocky planetary bodies which is summarized in Table 4. There can be no serious disputes about the timing of peak volcanism in different planetary bodies [ 12, 19-20] like Mars ( Tharsis volcanism) , Moon ( Mare volcanism ) and Venus ( period of Resurfacing).Using the LIP data base we have identified in a previous study [7] the number of volcanic eruptions in Earth with VEI ( 9-10) and VEI ( 10-11) etc for each 500 My period from 3 Gyrs to the present . This inferred mainly from area of volcanic eruptions and projected volume of such eruptions.The peak volcanism in Earth is inferred to occur between 0.5-1 Gyrs where we can find maximum occurrence of very intense LIP events ( VEI 10-11). It is interesting to find that inner core formation In Earth happened during the peak volcanism phase [14].

When dynamo in Mars was active around 4 Gyrs it generated moderate to strong magnetic fields and these events happened in association with peak phase of volcanism in this planet [19] Some models suggest onset of Plate tectonics during this period.[20-21] In this phase the CMB heat flux in Mars is inferred to be above the critical value required to operate the thermally drive dynamo [2]. During the declining phase of volcanism (roughly from 3.9 Gyrs to 3.6 Gyrs) the thermal dynamo is inferred to be active. CMB heat flux is critical in this phase and expected to be in the range 5-20 mW/m$^2$ as suggested by some previous studies [3, 22]. During the cessation phase of volcanism around 3.5 Gyrs or just after the thermal dynamo in Mars has either stopped working or has become weak.[11] In this phase CMB heat flux is expected go below the critical value.

From different studies we can find that the dynamo in Moon produced strong magnetic fields in the range 30-100 microtesla [8] during the period between 4.2 Gyrs to 3.6 Gyrs. We expect that Mare volcanism in Moon is possibly extended back to 4.2 Gyrs instead of 3.9 Gyrs as suggested in earlier studies. This is the period of peak volcanism in Moon where we can expect supercritical CMB flux if Moon had a thermally driven dynamo. The formation of core in Moon is inferred to be 4.1 Gyrs ago [23] and it justifies an active dynamo generating strong lunar magnetic fields from 4.2 Gyrs onwards. The declining phase of volcanism in Moon is inferred by us to be short between 3.6 to 3.3 Gyrs. This is associated with a sharp decrease in lunar magnetic fields inferred during this period. In this phase the CMB heat flux is decreasing but remained within the critical range 5-10 mW/m$^2$ as suggested by earlier studies[3.22]. The lunar dynamo either ceased or became very weak around 3.2 Gyrs which is just after the cessation phase of volcanism. In this phase the CMB heat flux is obviously sub-critical. A model of the relative change in CMB values in a magnetised rocky planet during different phases of major volcanism in that planet is shown in Fig 1.

The close associations found by us between volcanism and magnetic field evolution in Mars and Moon may be also true for Earth also. For the case of Earth during the peak phase of volcanism (0.5 -1 Gyrs) significant increase in internal magnetic fields is inferred which is possibly related to the formation of solid inner core during that period [24]. The critical CMB heat flow required for geodynamo to be active is inferred between 3-5 TW which is equivalent to a heat flux of about 6-10 mW/m$^2$[25-26] and is comparable to the critical CMB flux inferred for Moon [3] A recent study suggest a value of current CMB heat flow of 3.7-4.7 TW in Earth [27] which is close to the critical value required for the working of geo dynamo. If major volcanism in Earth ceases in near geological future it may shut down or significantly weaken the geodynamo which will be hazardous to life in Earth

The results in this paper suggest strong coupling between the dynamical phenomena happening in the core, mantle and crust in the rocky planetary objects in our solar system.The present findings be useful for modelling geological time evolution of volcanism and magnetism in rocky extrasolar planets.[28] The age of the extrasolar planet and its current phase of volcanism may be important for deciding the habitability conditions in these planets in addition to other criteria. We will address this problem in more detail in a future publication.

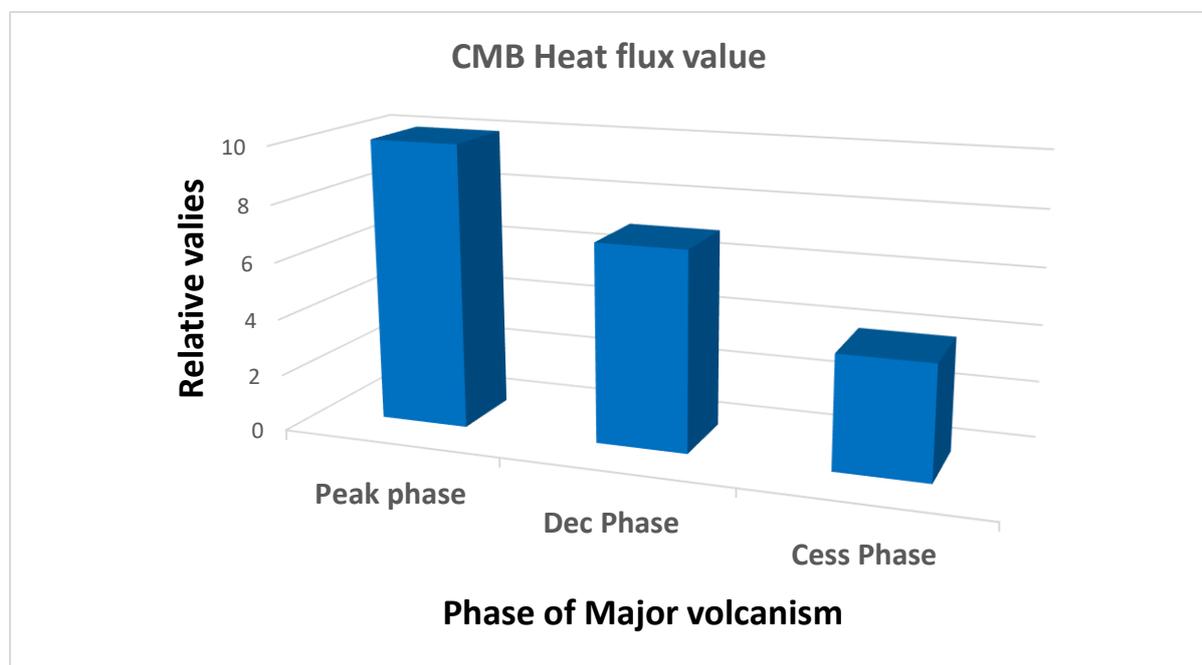

**Fig 1 A model for the relative values of Core to Mantle Boundary (CMB) Heat flux which controls the nature of planetary magnetic fields during (a) Peak phase (Super critical) (b) Declining phase (Critical) and (c) Cessation phase (Sub critical) of major volcanism in rocky planetary objects in our solar system**